\documentclass[aps,pra,twocolumn,amsmath,amssymb,nofootinbib,showpacs,superscriptaddress,longbibliography]{revtex4-1}
\usepackage[english]{babel}
\usepackage{latexsym}
\usepackage{graphics}
\usepackage{graphicx}
\usepackage{epsfig}
\usepackage{color}
\usepackage{bm}
\usepackage{amsmath}
\usepackage{amssymb}
\usepackage{amsthm}
\usepackage{dcolumn}
\usepackage{bm}
\usepackage{float}
\usepackage{hyperref}
\usepackage{color}
\usepackage{epstopdf}
\usepackage{braket}
\usepackage{tikz}
\usepackage{cleveref}
\usepackage{changes}

\theoremstyle{remark}

\begin{document}

\preprint{APS/123-QED}

\title{Simulating quantum circuits using \\ the multi-scale entanglement renormalization ansatz}

\author{A.V. Berezutskii}
\affiliation{Institut Quantique and D\'epartement de Physique, Universit\'e de Sherbrooke, Qu\'ebec, Canada J1K 2R1}

\author{I.A. Luchnikov}
\affiliation{Russian Quantum Center, Skolkovo, Moscow 143025, Russia}
\affiliation{National University of Science and Technology ``MISIS'', Moscow 119049, Russia}

\author{A.K. Fedorov}
\email{akf@rqc.ru}
\affiliation{Russian Quantum Center, Skolkovo, Moscow 143025, Russia}
\affiliation{National University of Science and Technology ``MISIS'', Moscow 119049, Russia}

\date{\today}
\begin{abstract}
Understanding the limiting capabilities of classical methods in simulating complex quantum systems is of paramount importance for quantum technologies. Although many advanced approaches have been proposed and recently used to challenge quantum advantage experiments, novel efficient methods for the approximate simulation of complex quantum systems are still in high demand. Here we propose a scalable technique for approximate simulations of intermediate-size quantum circuits on the basis of the multi-scale entanglement renormalization ansatz (MERA) and Riemannian optimization. The MERA is a tensor network, whose geometry together with orthogonality constraints imposed on its tensors allow approximating many-body quantum states lying beyond the area-law scaling of the entanglement entropy. We benchmark the proposed technique for brick-wall quantum circuits of up to 243 qubits with various depths up to 20 layers. Our approach paves a way to exploring efficient simulation techniques for quantum many-body systems.
\end{abstract}

\maketitle

\section{Introduction}

Recent progress in experiments with noisy intermediate-scale quantum (NISQ) devices on demonstrating advantages in solving specific problems, such as random circuit sampling~\cite{arute2019quantum, wu2021strong, zhu2021quantum} and boson sampling~\cite{zhong2020quantum, wang2019boson, zhong2021phase}, has stimulated a new wave of studies of the fundamental limits of classical techniques in the exact and approximate simulation of interacting many-body quantum systems. Specifically, various techniques using matrix product states (MPS)~\cite{vidal2003efficient, zhou2020limits}, projected entangled-pair states (PEPS)~\cite{guo2019general}, fractal states~\cite{Petrova2024}, neural network quantum states (NNQS)~\cite{jonsson2018neural, carrasquilla2021probabilistic, carleo2021classical} have been used in simulating quantum circuits of dozens of qubits. 
New tensor network contraction techniques have been successfully used for the efficient simulation of the quantum advantage circuits~\cite{pan2021simulating, pan2021solving, gray2021hyper, tindall2024efficient}. 
However, these advanced methods still suffer from many challenges. For example, MPS-based simulators are fundamentally restricted by the area-law scaling of the entanglement entropy. 
NNQS-based simulators suffer from inaccuracies induced by the noise coming from variational Monte Carlo (VMC)-based calculations~\cite{carleo2017solving, carleo2012localization}. The most advanced tensor network contraction algorithms for the classical simulation of quantum supremacy circuits provide only measurement samples and do not provide the entire final state.

One of the promising solutions for the problems listed above consists in building a simulator based on multi-scale entanglement renormalization (MERA)~\cite{vidal2007entanglement, vidal2009entanglement, evenbly2009algorithms}. Multi-scale entanglement renormalization is a powerful numerical technique mostly used to analyze critical behavior of one-dimensional quantum systems. It utilizes a tensor network known as the multi-scale entanglement renormalization ansatz to parametrize a many-body quantum state. In contrast to MPS tensor network, which only supports an area law of entanglement entropy scaling in one dimension (1D), i.e., $S_{\rm MPS}(L) \sim {\rm const}$, where $L$ denotes system size, MERA supports logarithmic entanglement entropy scaling with length of a $1D$ quantum many-body system, i.e., $S_{\rm MERA}(L) \sim \log(L)$~\cite{evenbly2011tensor}. Since this logarithmic behavior of the entanglement entropy perfectly matches the scaling of the entanglement entropy of ground states of quantum critical systems~\cite{evenbly2009entanglement, evenbly2011tensor}, MERA thus stands out as a perfect tool for studying quantum criticality. Besides quantum criticality, the improved entanglement entropy scaling is a remarkable feature that can be exploited to build better classical simulator of quantum computations. In contrast to neural-network approaches, MERA allows the exact calculation of expectation values in most cases as well as MPS; therefore, a MERA-based simulator does not suffer from the VMC-induced noise. Meanwhile, MERA can describe many-body quantum states beyond the area law of entanglement entropy scaling as well as neural networks. Moreover, one can tune MERA by using efficient Riemannian optimization methods with preconditioning~\cite{hauru2021riemannian, luchnikov2021riemannian, luchnikov2021qgopt} (also referred to as a quantum natural gradient~\cite{stokes2020quantum}). All these features make MERA a compromise solution in between MPS and NNQS that takes the best of both worlds.

In this paper, we propose and benchmark a MERA-based technique for the approximate simulation of arbitrary quantum circuits. Within our technique, we fix the structure of the MERA and update the ``elementary'' tensors of the MERA each time we apply a new quantum gate to the state parametrized by the MERA. Although this procedure is not always possible to do without approximation, we can find the closest MERA to the exact circuit state. At each step, the update of the MERA tensors is implemented through maximization of the gate fidelity, which formulates a corresponding optimization problem with orthogonality constraints~\cite{edelman1998geometry}. The significant simplification in this procedure comes from the fact that we update only blocks of the network that are not excluded because of the orthogonality constraints. We benchmark the proposed method for intermediate-size quantum circuits of up to 243 qubits with depths ranging from 4 to 20.

The paper is organized as follows. Section~\ref{sec:methods} provides a general description of the MERA architecture and the MERA-based quantum computations simulation protocol. Section~\ref{sec:comparison} is devoted to the comparison of the proposed protocol to an MPS-based one. Then, in Section~\ref{sec:results}, we present our simulation results of random quantum circuits. Finally, we summarize and conclude our results in Section~\ref{sec:discussion}.

\section{MERA tensor network}\label{sec:methods}

\begin{figure}[ht]
    \centering
    \includegraphics[scale=0.8]{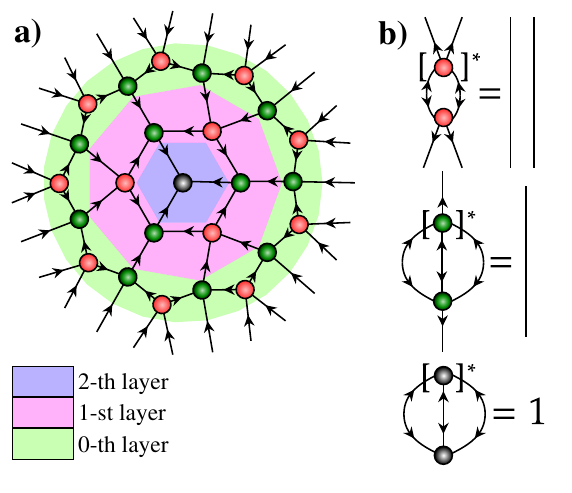}
    \caption{{\bf (a)} {\bf Diagrammatic representation of a three-layer ternary MERA tensor network.} The dangling edges represent the physical indices of a many-qubit quantum state $\ket{\psi}$. Red circles represent disentanglers and green circles represent isometries. The gray circle in the middle represents the top layer consisting of a single tensor that also can be viewed as an isometry which maps three input Hilbert spaces to a scalar. The arrows denote directions from the input Hilbert spaces to the output ones. {\bf (b)} {\bf Diagrammatic representation of the orthogonality properties, i.e., $vv^\dagger = I$, $uu^\dagger = u^\dagger u = I$ and $t^\dagger t = 1$, of elementary tensors.} The straight lines represent identity operators. The asterisk sign denotes complex conjugation.}
    \label{fig:mera_main}
\end{figure}

To build a quantum circuit simulator using the MERA tensor network, we begin by discussing the MERA itself, which serves as an ansatz for a many-qubit wave function $\ket{\psi}$. Without loss of generality, we choose a basic ternary MERA architecture~\cite{evenbly2009algorithms}. The three-layer ternary MERA is represented in Fig.~\ref{fig:mera_main} (a) in terms of the Penrose graphical notation~\cite{orus2014practical, bridgeman2017hand}.
The number of layers $M$ determines the number of subsystems (qubits) as $n = 3^M$. Therefore, the number of qubits in the ternary MERA represented in Fig.~\ref{fig:mera_main}-(a) is $n=27$. The MERA tensor network is built from two types of elementary tensors, namely isometries and disentanglers. One can think of an isometry as the following linear isometric map,
\begin{eqnarray}
    &&v: {\cal H}_1^{\rm in}\otimes {\cal H}_2^{\rm in} \otimes {\cal H}_3^{\rm in}\rightarrow {\cal H}^{\rm out},\nonumber\\
    && vv^\dagger = I,
\end{eqnarray}
where $\dagger$ stands for the Hermitian conjugation, $I$ is the identity operator, ${\cal H}^{\rm in}_i$ is the $i$-th input Hilbert space and ${\cal H}^{\rm out}$ is an output Hilbert space. The dimensions of the Hilbert spaces are connected as follows:
\begin{equation}
    {\rm dim}\left({\cal H}^{\rm out}\right) = 
    \begin{cases}
    \prod_{i=1}^3 {\rm dim}\left({\cal H}_i^{\rm in}\right),\text{ if }\prod_{i=1}^3 {\rm dim}\left({\cal H}_i^{\rm in}\right)\leq \chi,\\
    \chi,\text{ otherwise},
    \end{cases}
\end{equation}
where $\chi$ is the maximal allowed dimension of an index. This parameter determines the expressivity of a tensor network which is defined as the ability of the tensor network to represent quantum states with increasing complexity, quantified by the scaling of the maximum entanglement entropy. In addition, $\chi$ determines the numerical complexity of operations within the tensor network. A disentangler can be viewed as the following linear unitary map,
\begin{eqnarray}
    &&u: {\cal H}_1^{\rm in}\otimes {\cal H}_2^{\rm in} \rightarrow {\cal H}^{\rm out}_1 \otimes {\cal H}^{\rm out}_2,\nonumber\\
    && u^\dagger u = u u^\dagger = I,
\end{eqnarray}
where the dimensions of the input and output spaces are the same:
\begin{eqnarray}
    &&{\rm dim}\left({\cal H}_1^{\rm in}\right) = {\rm dim}\left({\cal H}_1^{\rm out}\right),\\
    &&{\rm dim}\left({\cal H}_2^{\rm in}\right) = {\rm dim}\left({\cal H}_2^{\rm out}\right).
\end{eqnarray}
The top layer of the MERA consists of only one tensor $t$ that can also be treated as the following isometric map:
\begin{eqnarray}
    &&t: {\cal H}_1^{\rm in}\otimes {\cal H}_2^{\rm in} \rightarrow \mathbb{C}, \nonumber \\
    && t^\dagger t = 1.
\end{eqnarray}
The diagrammatic interpretation of the orthogonality property of these tensors is given in Fig.~\ref{fig:mera_main} (b).

\begin{figure*}[ht]
  \includegraphics[scale=0.8]{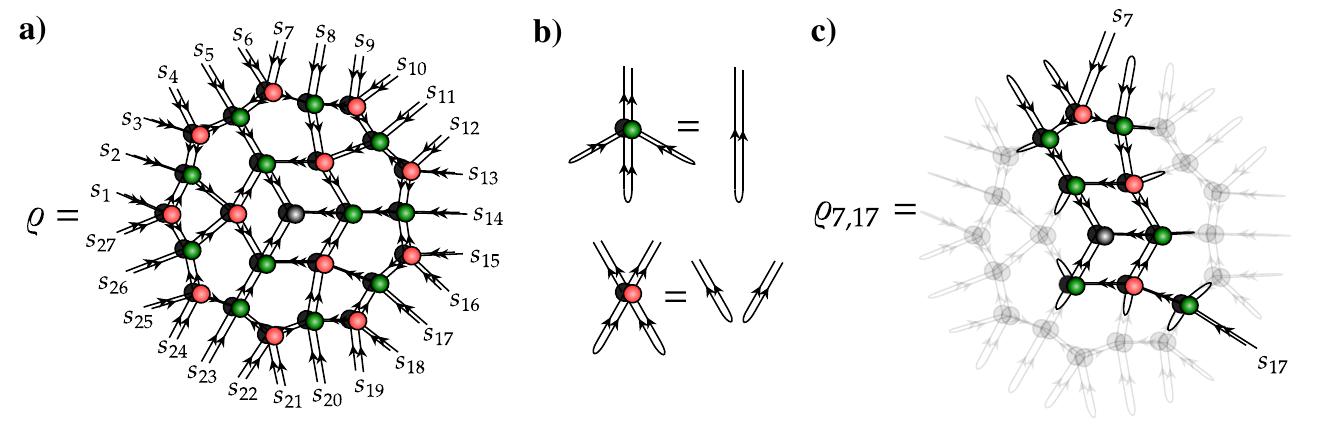}
  \caption{{\bf (a)} {\bf Diagrammatic representation of the density matrix $\varrho = \ket{\psi}\bra{\psi}$ of the entire system} built upon the MERA tensor network. The top colored layer and the underlying shadowed layer correspond to $\ket{\psi}$ and $\bra{\psi}$ accordingly while $s_i$ denotes an $i$-th subsystem. {\bf (b)} {\bf Reduction rules based upon the orthogonality property.} The implementation of these rules significantly reduces the computational complexity involved in evaluating reduced density matrices. {\bf (c)} {\bf The partial density matrix of the seventh and the $17$-th subsystems.} The U-shaped loops connecting the physical indices of $\ket{\psi}$ and $\bra{\psi}$ indicate a partial trace over the subsystem. Due to the orthogonality property only a part of the isometries and the disentanglers contributes to the resulting density matrix.}
  \label{fig:mera_partial}
\end{figure*}

The orthogonality property of elementary tensors substantially reduces the numerical complexity of operations with MERA. Let us demonstrate this by calculating a reduced density matrix of two qubits. The graphical representation of the entire system's density matrix built upon the three-layer ternary MERA is given in Fig.~\ref{fig:mera_partial}-(a).
It is represented as the two layers of tensor diagrams, i.e. the first layer represents $\ket{\psi}$ and the second layer represents $\bra{\psi}$ in $\varrho = \ket{\psi}\bra{\psi}$. To get a reduced density matrix, we start to take traces over the qubits that we want to exclude from consideration, i.e., we trace out the physical indices of MERA. Due to the orthogonality property one can introduce the reduction rules represented in Fig.~\ref{fig:mera_partial}-(b). After having applied these rules, we obtain the tensor network describing the reduced density matrix whose diagrammatic representation is given in Fig.~\ref{fig:mera_partial} (c). Noteworthy, this tensor network is defined by only a part of elementary tensors. Such subset of tensors is usually referred to as \emph{causal cone} \cite{evenbly2014scaling}. The asymptotic contraction complexity of the causal cone is $O(\chi^8)M= O(\chi^8)\log(n)$~\cite{evenbly2009algorithms}, which enjoys the logarithmic scaling with the number of qubits $n$.

\section{Application of Quantum Gates}

\begin{figure*}[ht]
  \includegraphics[scale=0.8]{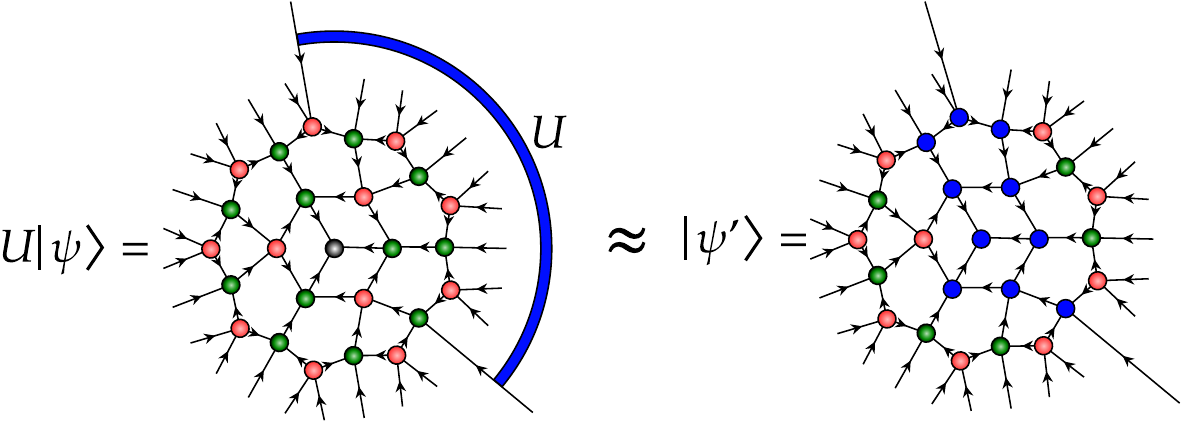}
  \caption{
  	{\bf Approximate equality between MERA with an attached two-qubit gate and MERA with updated elementary tensors in the corresponding causal cone.} The causal cone that is being updated is highlighted by the blue color.}
  \label{fig:mera_update}
\end{figure*}

\begin{figure}[ht]
    \centering
    \includegraphics[scale=0.8]{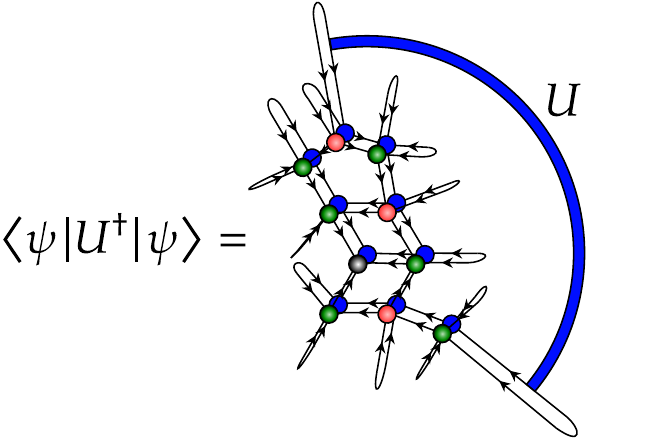}
    \caption{{\bf The diagrammatic representation of the overlap $\bra{\psi}U^\dagger\ket{\psi}$.} Only causal cones from $\ket{\psi'}$ and $\ket{\psi}$ survive due to the orthogonality property. The bottom layer represents tensors that are being updated from the causal cone of $\ket{\psi'}$, the top layer is the causal cone of $\ket{\psi}$.}
    \label{fig:mera_u_braket}
\end{figure}

To simulate quantum circuits using the MERA tensor network, quantum gates must be applied to the network. The application of a single-qubit gate is straightforward—one simply contracts the gate with the corresponding elementary tensor to which it is attached. The orthogonality property of the elementary tensor is preserved due to the gate's unitarity. However, the application of two-qubit gates is more complex. We employ the variational principle to update the elementary tensors within the causal cone, ensuring that the resulting MERA accurately approximates the state after the two-qubit gate is applied. In Fig.~\ref{fig:mera_update}, we give the diagrammatic interpretation of this update.
The variational principle is the maximization of the fidelity $F(\psi, \psi') = \left|\bra{\psi'} U \ket{\psi}\right|^2$ between the updated MERA $\ket{\psi'}$ and the MERA with the attached two-qubit gate $U \ket{\psi}$ with respect to the elementary tensors in the causal cone. For instance, an overlap $\bra{\psi} U^\dagger \ket{\psi}$ is given in Fig.~\ref{fig:mera_u_braket}, it involves only elementary tensors from the causal cone due to the orthogonality property.

We maximize $F(\psi, \psi')$ by using the preconditioned Riemannian Adam optimizer on the complex Stiefel manifold \cite{hauru2021riemannian, becigneul2018riemannian, li2020efficient} which preserves orthogonality constraints. This optimizer requires multiple calculations of the gradient of the fidelity and multiple calculations of the preconditioner which is properly defined in Ref.~\cite{hauru2021riemannian}. To calculate both the gradient and the preconditioner, we use the automatic differentiation technique~\cite{liao2019differentiable}. The complexity of the automatic differentiation-based computation of the gradient and preconditioners is the same as the complexity of the fidelity computation~\cite{griewank1989automatic, baur1983complexity}. Therefore, under the assumption of the same number of optimization steps for any $\chi$ and $n$, the complexity of the overall optimization routine scales as the complexity of the density matrix calculation, i.e., as $O(\chi^8)\log(n)$, though possibly with a large prefactor. For multiple quantum gates in a quantum circuit one needs to solve the optimization problem for each gate. Noteworthy, the technique is similar in spirit to the one presented in \cite{rizzi2008simulation}.

\begin{figure}[ht]
    \centering
    \includegraphics[scale=1.0]{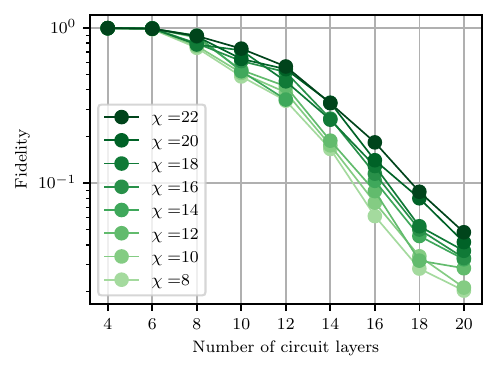}
    \caption{{\bf Fidelity estimation ${\cal F}$ for the MERA-based quantum circuit simulation.} The $x$-axis represents the depth $k$ of the brick-wall quantum circuits with $243$ qubits. Different curves correspond to different $\chi$ ranging from $8$ (the lightest) to $22$ (the darkest).}
    \label{fig:MERA-243-qubits-fidelity}
\end{figure}

\begin{figure}[ht]
    \centering
    \includegraphics[scale=1.0]{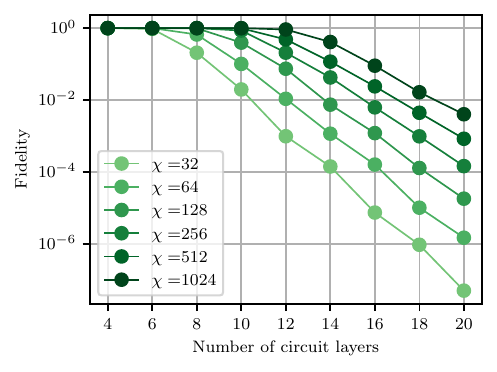}
    \caption{{\bf Fidelity dynamics in the $243$ qubits and $20$ layers circuit in the MPS-based simulation for different bond dimensions.} This is the analog of Fig.~\ref{fig:MERA-243-qubits-fidelity} but for MPS-based simulation.}
    \label{fig:MPS-243-qubits-fidelity}
\end{figure}

\begin{figure}[ht]
    \centering
    \includegraphics[scale=1.0]{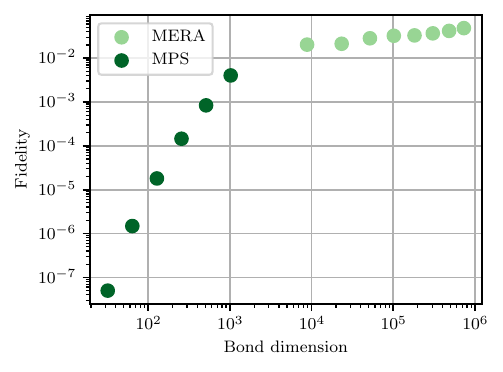}
    \caption{{\bf The final fidelity of the $243$ qubits and $20$ layers circuit simulation vs $\chi_{\rm mps}$ for MPS and MERA-based simulations.} One can see how MERA-based simulation results extend MPS-based results to much higher bond dimensions.}
    \label{fig:MPS-MERA-bonddim-comparison}
\end{figure}

\section{Numerical results of MERA-based simulation of random quantum circuits}\label{sec:results}

\begin{figure*}[ht]
    \centering
    \includegraphics{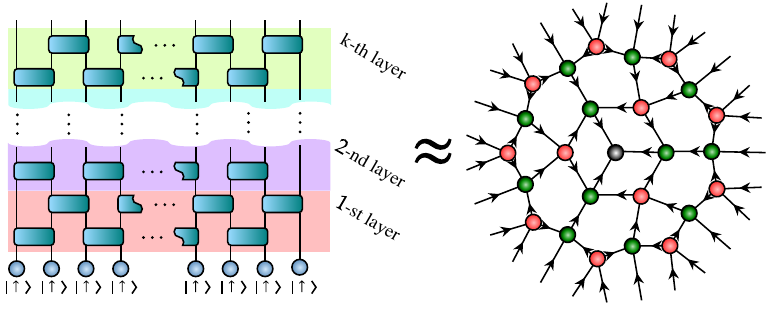}
    \caption{{\bf The entire circuit applied to the initial product state is being transformed to a MERA tensor network.} The left-hand side of the figure shows the tensor diagram of a brick-wall quantum circuit that is simulated via the proposed approach. The right-hand side of the figure shows the MERA tensor network after $nk$ updates which is approximately equal to the final state of the circuit.}
    \label{fig:MERA_to_circuit}
\end{figure*}

We test the proposed approach by simulating brick-wall quantum circuits with randomly sampled two-qubit gates. This is a standard benchmark used to evaluate the performance of a simulation approach~\cite{zhou2020limits} due to the high entanglement possessed by such circuits. The gates in the circuit are placed in a brick-wall order as shown in Fig.~\ref{fig:MERA_to_circuit}. Each gate is generated randomly in two steps: (i) one samples real and imaginary parts of $4\times 4$ matrix from the independent and identically distributed (i.i.d.) normal distribution ${\cal N}(0, I)$, where $I$ is the identity matrix; (ii) one performs QR decomposition of the sampled matrix and utilizes unitary matrix $Q$ as a two-qubit gate. Here, a QR decomposition of a matrix $A$ is expressed as $A = QR$, where $Q$ is an orthogonal matrix and $R$ is an upper triangular matrix.

The initial state is set to $\bigotimes_{i=0}^{n-1} \ket{\uparrow}$, where $\uparrow$ stands for the Bloch vector pointing to the north pole of the Bloch ball. In order to set the MERA tensor network to this initial state we maximize the sum of one-qubit fidelities $\sum_{i=1}^{n-1} \bra{\uparrow}\varrho_i \ket{\uparrow}$, where $\varrho_i$ is the $i$-th qubit density matrix computed from the MERA, using the same techniques that we use for a two-qubit gate application.

Having the MERA tensor network set in the initial state $\bigotimes_{i=0}^{n-1} \ket{\uparrow}$ one performs its elementary tensors update by maximizing fidelity for each subsequent gate. After finishing all the updates, one ends up with an approximation of the final quantum state.

In order to evaluate the accuracy of the obtained approximation, we calculate the following quantity:
\begin{equation}
    \label{eq:fidelity_surrogate}
    {\cal F} = \prod_{i=1}^{M}\left|\bra{\psi_i}U_i\ket{\psi_{i-1}}\right|^2,
\end{equation}
where $M$ is the total number of gates, $U_i$ is the $i$-th gate and $\ket{\psi_i}$ is the state represented by the MERA tensor network after $i$ gates are applied. We note that the factors $\bra{\psi_i}U_i\ket{\psi_{i-1}}$ in Eq.~\eqref{eq:fidelity_surrogate} come for free as every term in the product Eq.~\eqref{eq:fidelity_surrogate} is the optimal value of the fidelity found within the optimization process for the $i$-th gate. In Ref.~\cite{zhou2020limits} it has been shown that ${\cal F}$ is an accurate surrogate of the exact fidelity between the exact final state of a quantum program and an approximation. The value of ${\cal F}$ ranges in the interval $[0, 1]$, and higher ${\cal F}$ indicates a better approximation. The value of ${\cal F}$ is equal to $1$ if and only if the final state is exact. One can also estimate an average fidelity of each two-qubit gate as follows,
\begin{equation}
    f \approx {\cal F}^{\frac{1}{M}}
\end{equation}
The value $\epsilon = 1 - f$ can be considered as the error rate per two-qubit gate or two-qubit gate infidelity frequently used to benchmark NISQ devices.

For the number of qubits $n=243$ corresponding to a five-layer MERA and $\chi$ ranging from $8$ to $22$, we have performed a MERA-based simulation of random brick-wall quantum circuits. We present ${\cal F}$ as a function of the number of layers $k$ in a quantum circuit and value of $\chi$ in Fig.~\ref{fig:MERA-243-qubits-fidelity}.

As expected, the accuracy of the simulation improves with increasing $\chi$, and deteriorates with the increase of the number of layers. The average fidelity of a two-qubit operation $f$ for $\chi=22$ is equal to $0.998$ and the corresponding per-gate error rate is $\epsilon = 0.002$ that are comparable with the fidelity and error rate of a two-qubit operation of real-world implementations of a quantum computers~\cite{arute2019quantum}.

\section{Comparison with the MPS-based simulation}\label{sec:comparison}

\begin{figure}[ht]
    \centering
    \includegraphics[width=\linewidth]{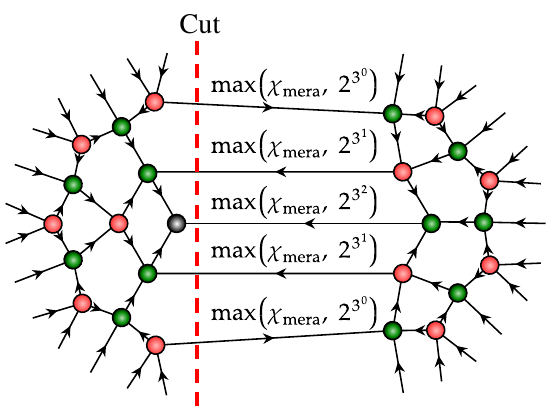}
    \caption{{\bf Visual interpretation of Eq.~\eqref{eq:mera_expressivity}.} Each intersected index has dimension at most $\chi_{\rm mera}$, which is why one takes the maximum of $\chi_{\rm mera}$ and $2^{3^l}$ which is the law of how the dimension of indices grows towards the top layer before reaching $\chi_{\rm mera}$.}
    \label{fig:mera_cut}
\end{figure}

In this section, we compare the quantum circuit simulations conducted using the MERA in Section~\ref{sec:results} with those performed using the MPS on the same circuits. We begin by discussing the expressivity of the MERA and the MPS, which is essential for the comparison. The expressivity of a tensor-network ansatz is commonly measured by its bond dimension, $\chi$, which sets an upper bound on the maximal entanglement entropy achievable within the MPS by $\log(\chi)$. To effectively compare the expressivities of MERA and MPS, it is thus necessary to relate their bond dimensions, $\chi_{\rm mps}$ and $\chi_{\rm mera}$. In order to perform this comparison, one needs to transform the MERA into the MPS. This transformation is always possible by simply contracting the tensors along the depth of the MERA. The bond dimension of the resulting MPS can be estimated in two steps: (i) First, cut the MERA into two approximately equal parts; (ii) then, compute the product of the dimensions of the indices intersected by the cut---the result gives an estimate of the MPS bond dimension. Consequently, the bond dimension of the resulting MPS can be approximated as follows:

\begin{eqnarray}
    \label{eq:mera_expressivity}
    \chi_{\rm mps} &&= \left[\prod_{l=0}^{\log_3(n) - 2}\max\left(\chi_{\rm mera}, \ 2^{3^l}\right)\right]^2,\nonumber\\ &&\times\max\left(\chi_{\rm mera}, \ 2^{\frac{n}{3}}\right),
\end{eqnarray}
where $l$ enumerates layers in MERA staring from $0$. See Fig.~\ref{fig:mera_cut} for visual interpretation of Eq.~\eqref{eq:mera_expressivity}.

Next, we perform the simulation from Section~\ref{sec:results} using the MPS with different bond dimensions. The results of the simulation are given in Fig.~\ref{fig:MPS-243-qubits-fidelity} which is the analog of the Fig.~\ref{fig:MERA-243-qubits-fidelity} for the MERA-based simulation. We compare the final layer fidelity of the MPS-based simulation with the MERA-based simulation and also compare $\chi_{\rm mps}$ which is computed for MERA using Eq.~\eqref{eq:mera_expressivity}. The results of the comparison are given in Fig.~\ref{fig:MPS-MERA-bonddim-comparison}.

It is known, that as the number of qubits in the system increases, both the complexity and expressivity of the MERA-based protocol increase logarithmically~\cite{evenbly2009entanglement, evenbly2011tensor, evenbly2009algorithms}. This scaling is unattainable for the MPS-based protocol, where complexity escalates significantly faster than expressivity in any scenario. Consequently, the MERA-based protocol offers an alternative for systems with a large number of qubits compared to MPS-based protocols which can be seen by comparing the fidelity plots in Figs.~\ref{fig:MPS-243-qubits-fidelity} and \ref{fig:MERA-243-qubits-fidelity} as well as in Fig.~\ref{fig:MPS-MERA-bonddim-comparison} which compares the fidelity versus the bond dimension. These figures further illustrate that MPS fidelity increases more rapidly with bond dimension, albeit starting from a lower baseline. Collectively, these observations provide a practical basis for selecting the most suitable protocol for particular simulation needs.


\section{Data availability}

The code used in the study is available from the corresponding author upon reasonable request. The MERA simulations were performed using QGOpt~\cite{luchnikov2021qgopt} while the MPS simulations were performed using mdopt~\cite{mdopt2022}.

\section{Discussion and outlook}\label{sec:discussion}

We have presented the protocol for the MERA-based classical simulation of arbitrary quantum circuits. We demonstrated that our approach can be used for the successful simulation of intermediate-size quantum circuits. Specifically, we used random quantum circuits bookmarking for 243-qubit checkerboard circuits. We have also discussed estimations showing that the MERA tends to outperform MPS-based simulators for large $n$.

An important next step could be extending this simulation to deeper MERA tensor networks and larger quantum circuits. Additionally, it is crucial to develop an efficient algorithm for sampling measurement outcomes from MERA tensor networks to more accurately replicate the behavior of real quantum computers.

\section*{Acknowledgments}

The work was supported by the RSF Grant No. 19-71-10092 (Section~\ref{sec:methods}; analysis of the MERA ansatz), Leading Research Center on Quantum Computing (Agreement No. 014/20; Section~\ref{sec:results}, development of the simulation method), and Russian Roadmap on Quantum Computing (experiments with intermediate-size circuits). A.B. acknowledges Calcul Qu\'ebec and Compute Canada for computing resources.

\bibliography{bibliography.bib}

\end{document}